\documentclass[12pt]{article}
\pdfoutput=1

\usepackage{amsmath,amssymb,amscd}
\usepackage{listings}
\usepackage{caption}
\usepackage{dsfont}
\usepackage{slashed}
\usepackage{color}

\usepackage[pdftex]{graphicx}
\usepackage{epstopdf}
\usepackage{subfigure}
\usepackage{epsfig}
\usepackage{listings}
\usepackage{caption}
\usepackage{cite}

\usepackage{multirow}

\newcommand{\vol}[1]{{\bf #1}}

\newcommand{\bea}{\begin{align}}
\newcommand{\eea}{\end{align}}
\newcommand{\beq}{\begin{equation}}
\newcommand{\eeq}{\end{equation}}
\newcommand{\nbea}{\begin{align*}}
\newcommand{\neea}{\end{align*}}
\newcommand{\nbeq}{\begin{equation*}}
\newcommand{\neeq}{\end{equation*}}

 \newcommand{\ssum}{\displaystyle\sum}
 
 \newcommand{\column}[1]{\left(\begin{array}{c} #1 \end{array}\right) }

\numberwithin{equation}{section}

\begin{document}



\baselineskip=21pt
\rightline{KCL-PH-TH/2013-35, LCTS/2013-23}
\vskip 1in

\begin{center}

{\large {\bf Enhancement of Majorana Dark Matter Annihilation Through Higgs Bremsstrahlung}}

\vskip 0.2in

{\bf Feng~Luo}\footnote{feng.luo@kcl.ac.uk}
and {\bf Tevong~You}\footnote{tevong.you@kcl.ac.uk}

\vskip 0.2in

{\small {\it

{Theoretical Particle Physics and Cosmology Group, Physics Department, \\
King's College London, London WC2R 2LS, UK}\\

}}

\vskip 0.2in

{\bf Abstract}

\end{center}

\baselineskip=18pt \noindent


{\small

For Majorana dark matter, gauge boson bremsstrahlung plays an important role in enhancing an otherwise helicity-suppressed s-wave annihilation cross-section. This is well known for processes involving a radiated photon or gluon together with a Standard Model fermion-antifermion pair, and the case of massive electroweak gauge bosons has also recently been studied. Here we show that internal Higgs bremsstrahlung also lifts helicity suppression and could be the dominant contribution to the annihilation rate in the late Universe for dark matter masses below $\sim 1$ TeV. Using a toy model of leptophilic dark matter, we calculate the annihilation cross-section into a lepton-antilepton pair with a Higgs boson and investigate the energy spectra of the final stable particles at the annihilation point. 
}




\section{Introduction}


As the latest Planck results indicate that dark matter (DM) forms $\sim26\%$ of the energy density of our Universe (in standard $\Lambda$CDM cosmology)~\cite{planck}, a new generation of upcoming experiments raises the prospects of elucidating its nature. Together with the discovery of a Higgs boson~\cite{discovery} and the direct exploration of the TeV scale at the LHC, the phenomenological window begins to narrow down the landscape of possibilities. Many well-motivated models of new physics provide DM candidates which may be observable through their annihilation with each other into Standard Model (SM) particles. Interpreting DM indirect detection experiments relies upon understanding the production of SM particles at the annihilation point, before they get propagated through astrophysical models to yield the final flux measured at Earth. It is thus essential to include all relevant processes when calculating DM self-annihilation rates. 

The velocity-weighted annihilation cross-section may be decomposed when off-resonance\cite{offresonance} into a velocity-independent $s$-wave part and a velocity-dependent $p$-wave part to order $v^2$ in the DM velocity, $\sigma v = a + bv^2 + \mathcal{O}(v^4)$. If the DM particle, $\chi$, is a Majorana fermion then the $s$-wave contribution of the two-to-two annihilation into a SM fermion-antifermion pair, $\chi\chi \to f\bar{f}$, is suppressed by $(m_f / m_\chi)^2$ and vanishes in the chiral limit $m_f \to 0$. The surviving $p$-wave contribution is itself velocity-suppressed, since for our current Universe $v\sim10^{-3} c$ in the Galactic halo. It was pointed out early on~\cite{originalbremsstrahlung} that photon and gluon bremsstrahlung corrections in $\chi\chi \to \gamma f \bar{f}$ and $\chi\chi \to g f \bar{f}$ processes lift the helicity suppression in Majorana dark matter annihilations. Despite these higher-order processes being reduced by an extra coupling and phase-space factor $\sim \alpha_{\text{em},s} / \pi$, the additional $s$-wave contribution is not velocity-suppressed and could therefore enhance the annihilation rate. 

In the last few years there has been renewed interest in bremsstrahlung corrections, this time with a massive electroweak gauge boson in the three-body final state\footnote{This contribution from internal bremsstrahlung in the hard process is to be distinguished from the soft collinear radiation off on-shell final-state fermions, which is logarithmically enhanced~\cite{ciafalonietal1}.}\cite{yaguna,belletal1,ciafalonietal2,garny1,ciafalonietal3, bargeretal, garny2, ciafalonietal4}. The annihilation rate in this case can also be larger than the helicity- and velocity-suppressed two-body process, with the subsequent decays of the $W^\pm/Z$ bosons phenomenologically relevant for the flux of antiprotons, neutrinos, photons and positrons measured on Earth. This effect is relevant for models that seek to explain the PAMELA~\cite{pamela2008,pamela2013} and AMS-02~\cite{ams02} positron excess without affecting the antiproton flux~\cite{pamelaantiproton} that is compatible with the expected astrophysical background \cite{ciafalonietal1}. The fragmentation products of electroweak gauge bosons open up the hadronic final state for leptophilic DM models and therefore place more stringent constraints. The impact on neutrino signatures from DM annihilation in the Sun has also been investigated \cite{neutrinofromsun}. 

The recent discovery of a Higgs boson turns this last theoretical piece of the SM jigsaw into experimental fact. Its phenomenological consequences in particle physics (and ``in space!"~\cite{higgsinspace}) can now be assessed more accurately. For the case of Majorana DM we find that the two-to-three $\chi\chi \to Hf\bar{f}$ process with a radiated Higgs also opens up the $s$-wave and can even be the dominant channel for $m_\chi \lesssim 1$ TeV. The purpose of this paper is to present a first calculation of the effects of Higgs bremsstrahlung in Majorana DM annihilation, using the toy model described in Section \ref{sec:darkmattermodel}. In Section \ref{sec:higgsstrahlung} the cross-section of this new Higgs-strahlung process is analysed and compared to that of the radiated $W^\pm, Z$ and $\gamma$ vector boson case. The subsequent decay of the Higgs and its effect on the flux of stable SM particles is considered in Section \ref{sec:energyspectra}. We conclude in Section \ref{sec:conclusion} with some comments on the importance of this effect for indirect detection experiments. Details of the Higgs-strahlung calculations and analytical expressions can be found in Appendix \ref{sec:calculation}.

\section{Dark Matter Model}
\label{sec:darkmattermodel}

We consider a Majorana fermion $\chi$, neutral under the SM gauge group, as the DM particle. $\chi$ is taken to be odd under an exactly conserved $Z_2$ symmetry, with SM particles being even, to ensure DM stability. With only this additional particle there are no dimension-four Lorentz- and gauge-invariant interaction terms with SM fermions.
This suggests either an effective Lagrangian approach~\cite{effectiveoperators} or adopting a minimal completion. We choose the latter option so as to include non-decoupled scenarios where the effective approach breaks down, and add an $\text{SU}(2)_L$ doublet scalar $\eta = \left( \eta^+ , \eta^0  \right)^T$ which is $Z_2$-odd, singlet under $\text{SU}(3)_c$ with hypercharge $1/2$ and mass $m_{\eta^\pm}, m_{\eta^0} > m_\chi$. 
We consider only the DM coupling to the first generation of leptons, treated as massless, by giving the $\eta$ doublet fields an electron lepton number of $-1$. The resulting Lagrangian is~\cite{ma,garny1}
\begin{equation}
\mathcal{L} = \mathcal{L}_\text{SM} + \frac{1}{2} {\bar \chi} i\slashed{\partial}\chi - \frac{1}{2}m_\chi{\bar \chi}\chi + (D_\mu \eta)^\dagger (D^\mu \eta) + [y_\text{DM} {\bar \chi}(L i\sigma_2 \eta) + \text{h.c.}]  - V_\text{scalar}	\, ,
\label{eq:lagrangian}
\end{equation}
where $L=\column{ {\nu_e}_L \\ e_L}$ and the scalar potential, including the SM Higgs doublet $\Phi = \column{ \phi^+ \\ \phi^0}$, can be written as
\begin{equation}
V_\text{scalar} = \mu_1^2 \Phi^\dagger \Phi + \frac{1}{2}\lambda_1(\Phi^\dagger\Phi)^2 + \mu_2^2 \eta^\dagger\eta + \frac{1}{2}\lambda_2(\eta^\dagger\eta)^2 + \lambda_D(\Phi^\dagger\Phi)(\eta^\dagger\eta) + \lambda_F(\Phi^\dagger\eta)(\eta^\dagger\Phi)	\, .
\label{eq:potential}
\end{equation}

For all values of $\mu_1^2$ and $\mu_2^2$, the condition that $V_\text{scalar}$ be bounded from below requires $\lambda_1 > 0$, $\lambda_2 > 0$, $\lambda_D > - \sqrt{\lambda_1 \lambda_2}$ and $\lambda_D + \lambda_F > - \sqrt{\lambda_1 \lambda_2}$~\cite{ma_boundbelow}. By assuming $\mu_1^2 < 0$ and $\mu_2^2 > 0$, the minimization of $V_\text{scalar}$ leads to a vacuum expectation value for only the $\phi^0$ field, $\langle \phi^0 \rangle = \sqrt{-\mu_1^2 / \lambda_1} \equiv v_{\text{EW}} \approx 174$ GeV. 
The physical masses of the scalar particles are then 
\begin{equation*}
m_H^2 = 2\lambda_1 v_\text{EW}^2		\, , \,  
m_{\eta^0}^2 = \mu_2^2 + (\lambda_D + \lambda_F) v_\text{EW}^2		\, , \,  
m_{\eta^\pm}^2 = \mu_2^2 + \lambda_D v_\text{EW}^2	\, .
\end{equation*}
We assume a SM Higgs with mass $\sim 125$ GeV throughout, consistent with the measured properties of the newly-discovered boson.
Note that $\lambda_F$ parametrizes the mass degeneracy between the charged and neutral $\eta$ scalars. We define the dimensionless ratios 
\begin{equation*}
r_{\pm,0} = \left(\frac{m_{\eta^\pm,\eta^0}}{m_\chi}\right)^2	\, ,
\end{equation*} 
though in practise we will specify $\lambda_F$ and $r_{\pm}$ with the neutral scalar mass fixed by the relation $r_0 = r_\pm + \lambda_F \frac{v_\text{EW}^2}{m_\chi^2}$.

Such a model is equivalent to a pure Bino DM interacting via an $\text{SU}(2)_L$ sfermion doublet in the minimal supersymmetric extension of the Standard Model (MSSM), and may be extended to encompass fully realistic theories. For example the case of a general neutralino DM has been calculated in full for electroweak gauge boson bremsstrahlung~\cite{neutralino}. As our aim is to illustrate the relative importance of Higgs bremsstrahlung it is not necessary to go beyond the simplified setup used here and widely elsewhere in the literature.

We see that the $\lambda_D$ and $\lambda_F$ terms have a similar form to the D-term and F-term, respectively, in the MSSM Lagrangian, where the $\eta$'s would then be the first generation left-handed selectron and sneutrino. In this scenario $\lambda_F$ is proportional to the square of the Yukawa coupling which vanishes in the chiral limit, while $\lambda_D$ is proportional to the square of the electroweak gauge coupling. However we note that in the MSSM the D-term Lagrangian is different for the left-handed selectron and sneutrino, since aside from the common $\text{U(1)}_Y$ coupling they also have different $\text{SU(2)}_L$ couplings.

In addition to the doublet model we will bear in mind the singlet model~\cite{Garny_singlet} in which the scalar $\eta$ is an $\text{SU(2)}_L$ singlet with hypercharge 1. The Lagrangian is identical to Eq.~(\ref{eq:lagrangian}) with $\lambda_F=0$ and the replacement $Li\sigma_2 \to e_R$ (as well as the appropriate gauge-covariant derivative for the scalar kinetic term). The singlet model corresponds to a Bino DM interacting with a right-handed slepton in the MSSM, and it is interesting to note that indeed in the stau-neutralino coannihilation region of the Constrained MSSM (CMSSM) the neutralino is mostly a Bino and the stau is mostly right-handed~\cite{sato}.

\section{Lifting Helicity Suppression With Higgs Bremsstrahlung}
\label{sec:higgsstrahlung}

\begin{figure}[h!]
\vspace*{-0.3in}
\begin{minipage}{8in}
\centering
\hspace*{-1.75in}
\includegraphics[height=4in,width=2.8in]{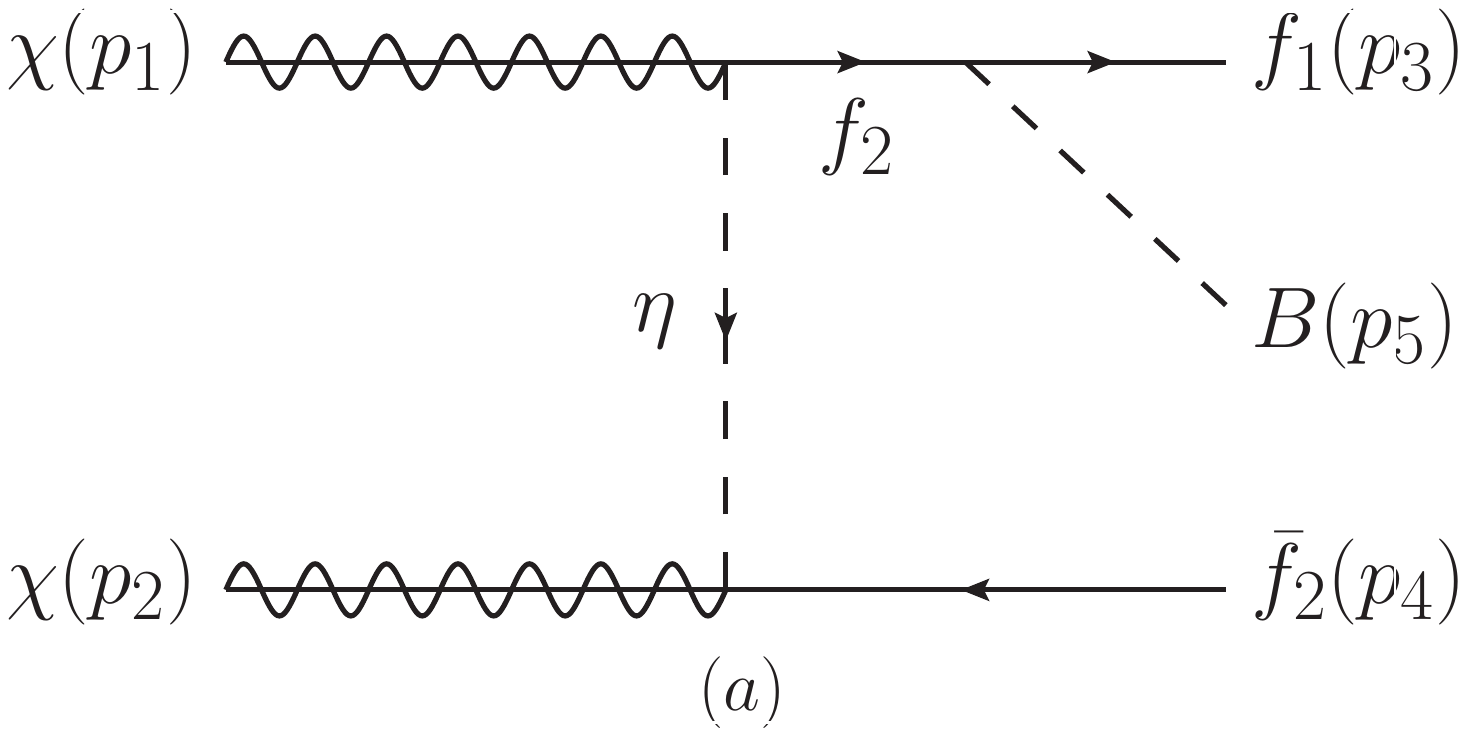}
\hspace*{-0.80in}
\includegraphics[height=4in,width=2.8in]{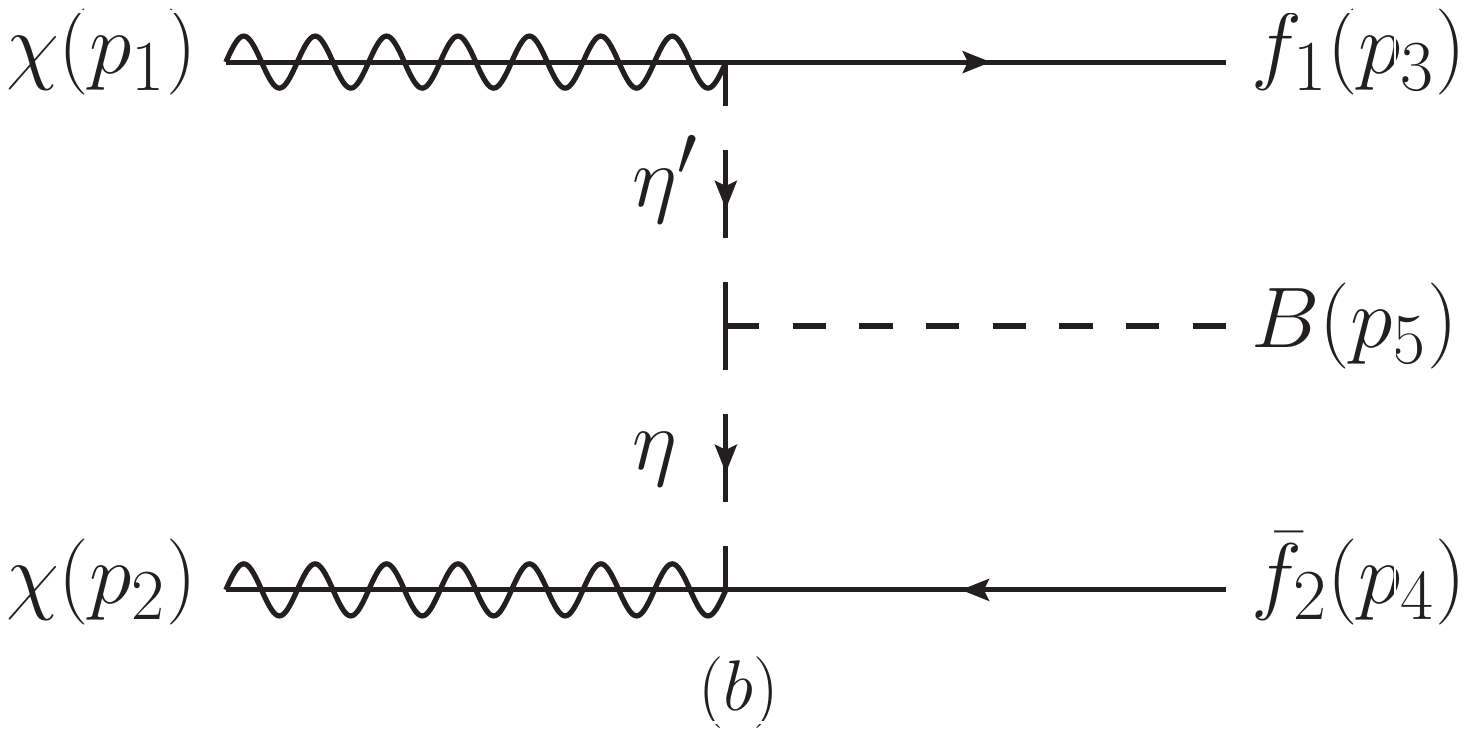}
\hspace*{-0.80in}
\includegraphics[height=4in,width=2.8in]{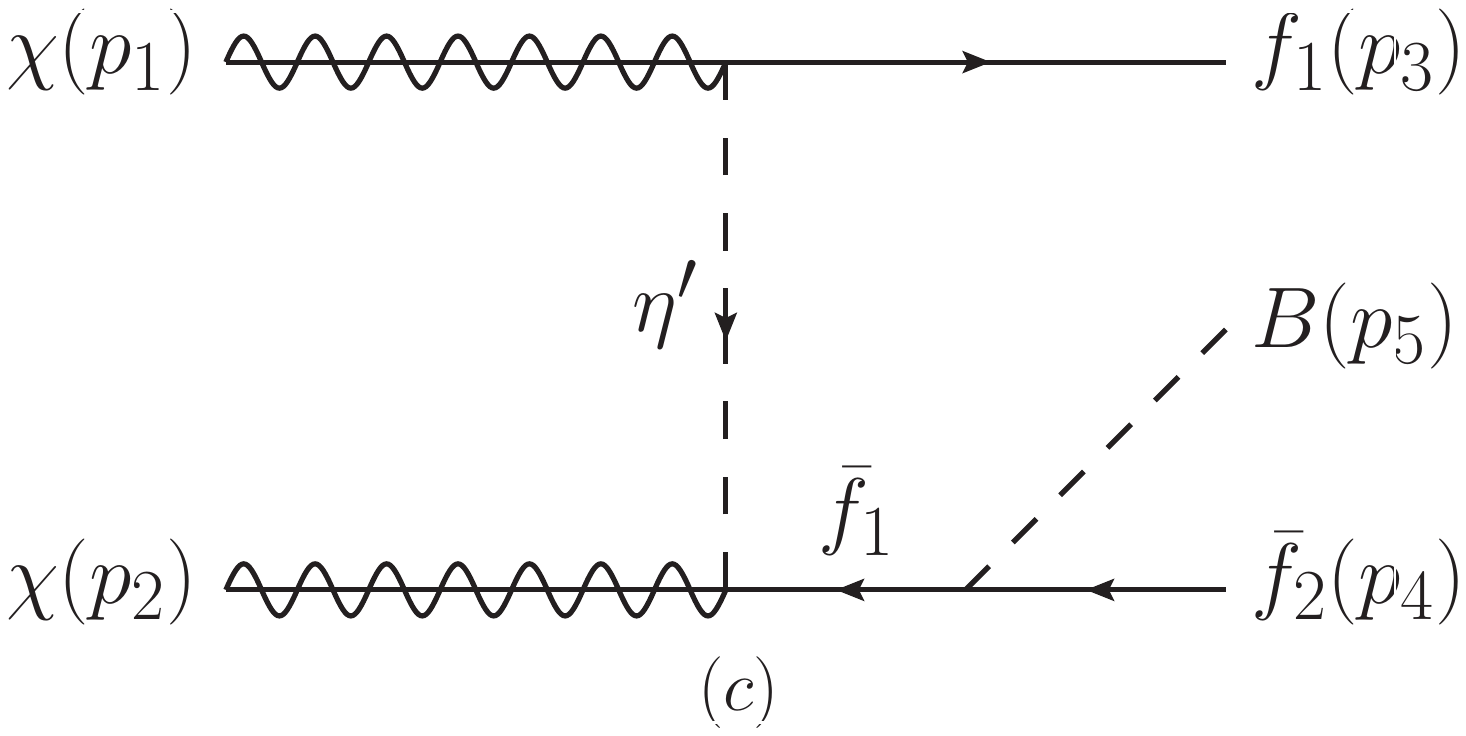}\\
\vspace*{-2.7in}
\hspace*{-1.75in}
\includegraphics[height=4in,width=2.8in]{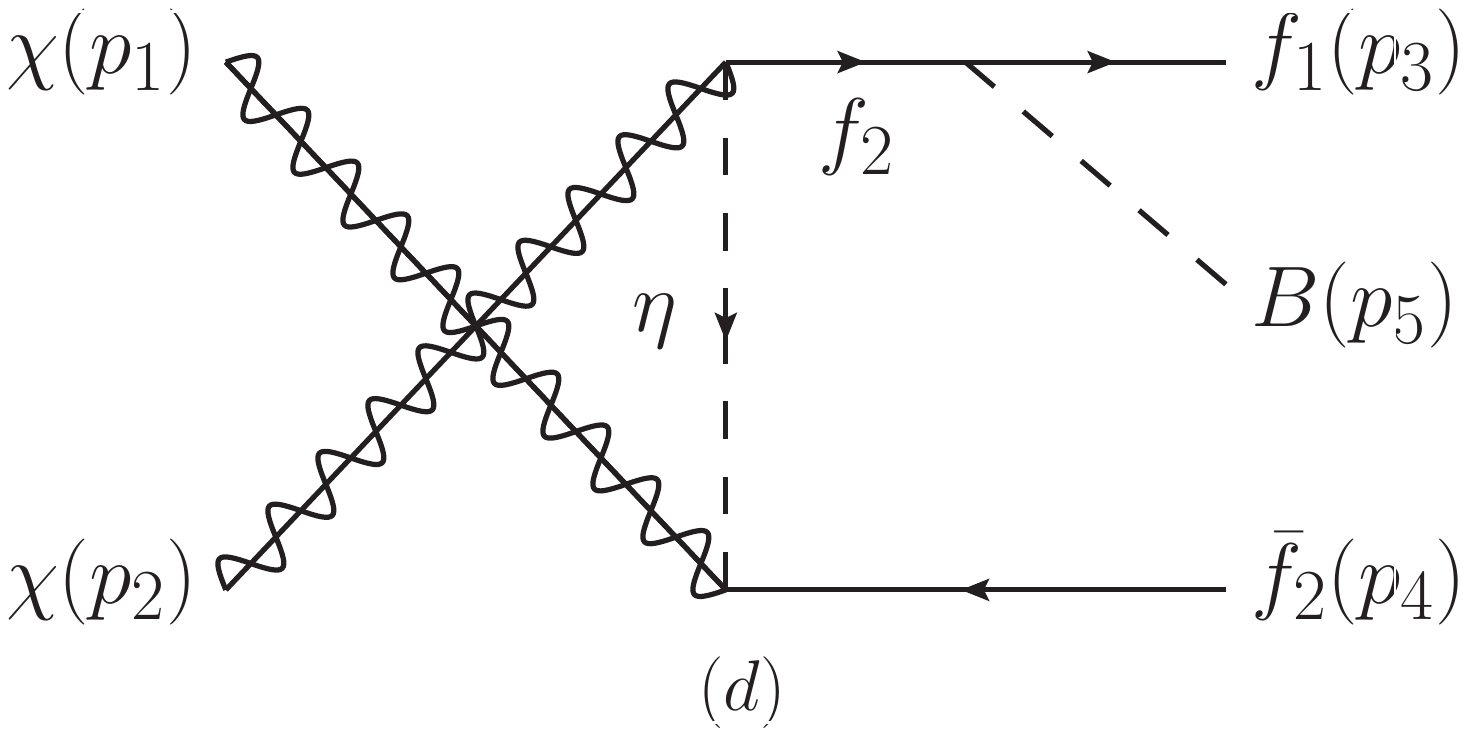}
\hspace*{-0.80in}
\includegraphics[height=4in,width=2.8in]{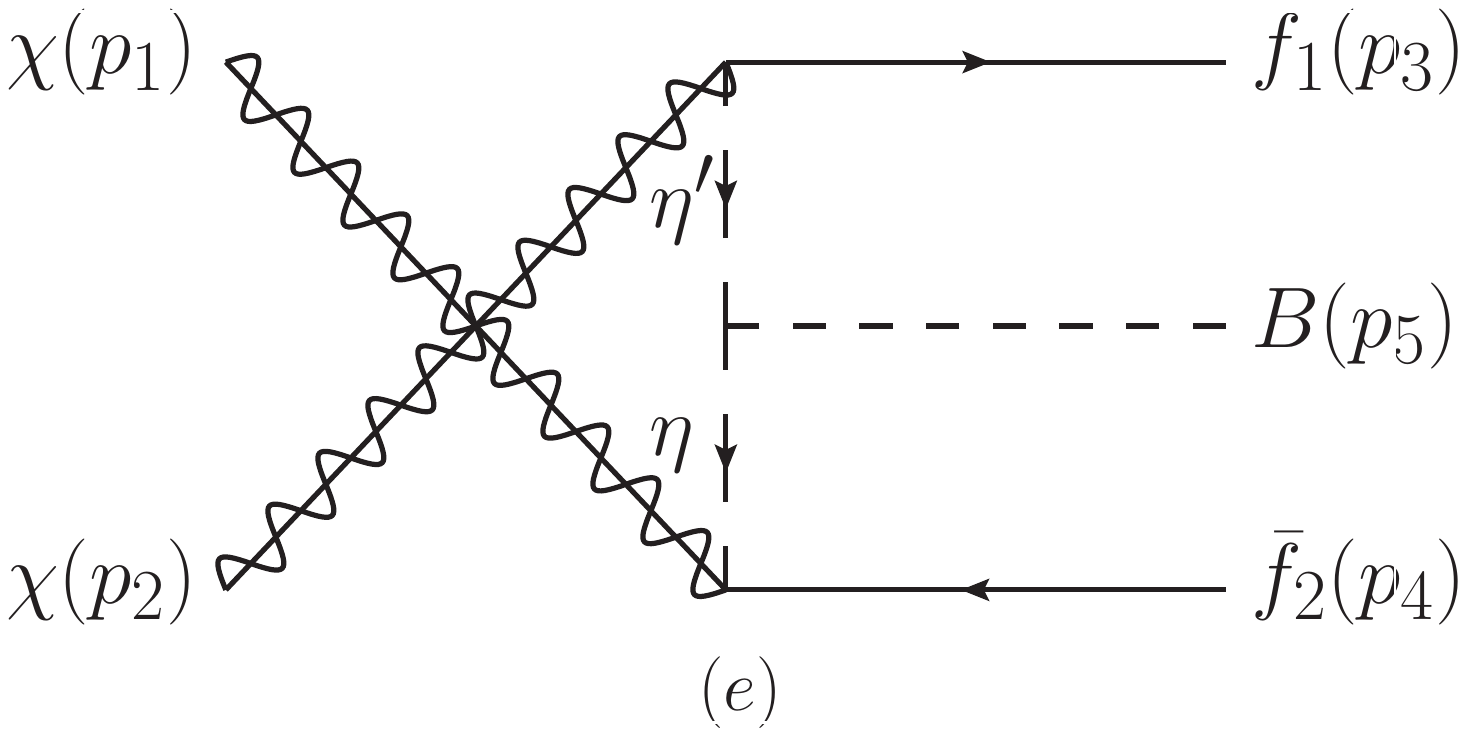}
\hspace*{-0.80in}
\includegraphics[height=4in,width=2.8in]{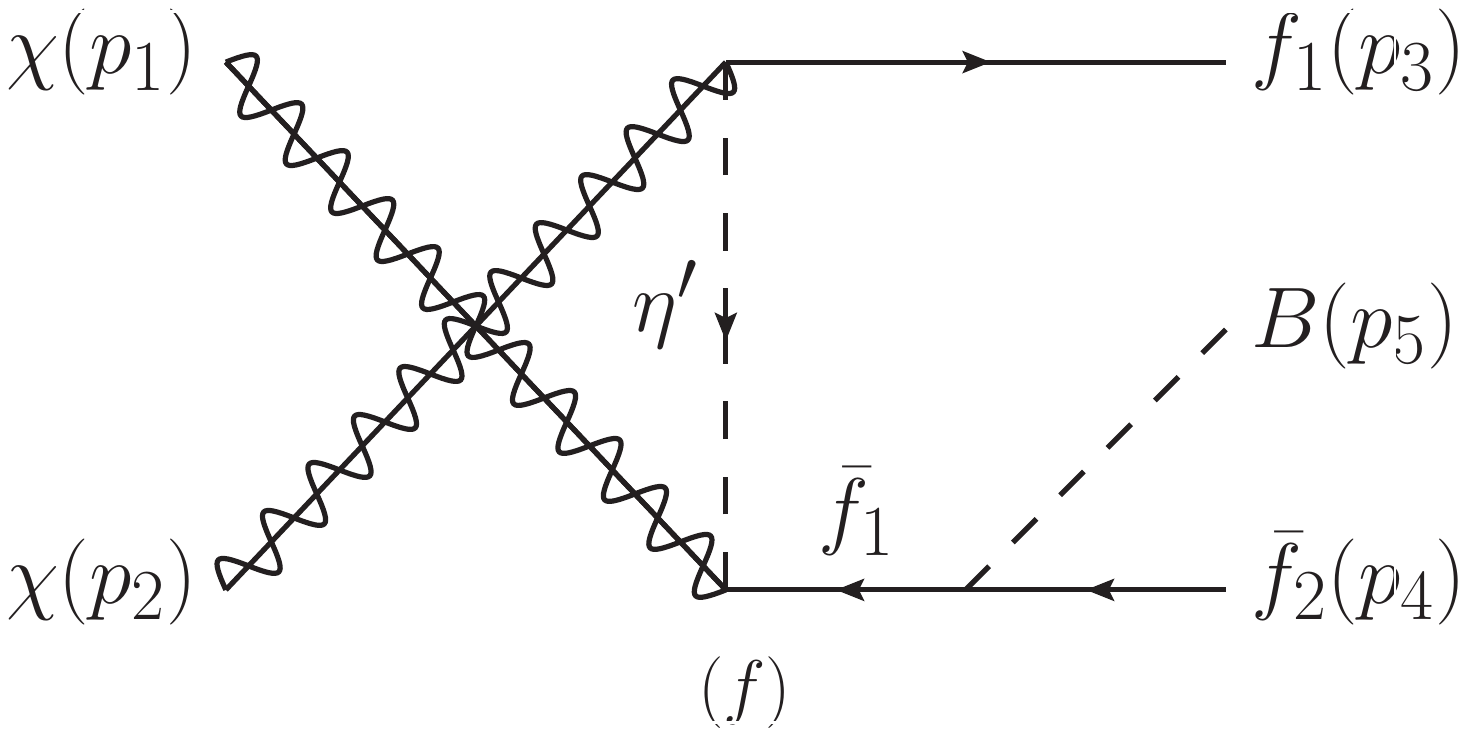}
\end{minipage}
\vspace*{-2.4in}
\caption{
{\it Generic Feynman diagrams for Majorana DM $\chi$ annihilating to SM fermions $e^\pm, \nu_e, \bar{\nu}_e$, with a boson $B = W^\pm, Z, \gamma, H$ in the three-body final state. The interactions are mediated by charged or neutral scalars $\eta, \eta^\prime$. When $B$ is electrically neutral, $f_1 = f_2$ and $\eta = \eta^\prime$.
}}
\label{fig:feynmandiagrams}
\end{figure}

The cross-section for the two-body $\chi\chi \to e^+ e^-, \nu_e \bar{\nu}_e$ process with massless final state fermions is easily found to be
\begin{equation*}
v\sigma |_{\chi\chi \to e^+ e^-, \nu \bar{\nu}} = \frac{y_{DM}^4(1+r_{\pm,0}^2)}{24\pi m_\chi^2(1+r_{\pm,0})^4} v^2 + \mathcal{O}(v^4)	\, , 
\end{equation*}
which contains no $s$-wave part. Including a gauge boson $\gamma,Z$ or $W^\pm$ in the final state adds the Feynman diagrams\footnote{These diagrams were created using JaxoDraw~\protect\cite{jaxodraw}. } shown in Fig.~\ref{fig:feynmandiagrams} to the annihilation cross-section, which are known to include an unsuppressed $s$-wave contribution. We calculate these using FeynCalc~\cite{feyncalc} with the method and analytical expressions summarised in Appendix \ref{sec:calculation}.

We briefly recall here why the two-body process turns out to be helicity suppressed by considering the wavefunction of the Majorana DM pair\footnote{See Ref.~\cite{2to2analysis} for a detailed analysis of the $2 \to 2$ case.}, which must be totally anti-symmetric for identical fermions. This means a symmetric (anti-symmetric) spin state requires an anti-symmetric (symmetric) spatial wavefunction, and the partial wave expansion tells us that these wavefunctions can only be expanded in the spherical harmonics denoted by odd (even) orbital angular momentum $l$. The velocity-unsuppressed $l=0$ partial wave must then be accompanied by an anti-symmetric spin state, which is the singlet fermion pair with total spin $S=0$ and $CP = (-)^{S+1}= -1$. If $CP$ is conserved then the total spin must also be zero in the final state, but this is not possible if the lepton and antilepton are massless since they are produced back-to-back with opposite momentum and must therefore have the same helicity. The addition of a lepton mass term provides the needed helicity flip, albeit suppressed by $(m_f / m_\chi)^2$.  An unsuppressed $s$-wave can be obtained by the addition of a vector boson in the final state, which allows a left-handed lepton to be produced with a right-handed antilepton while conserving total angular momentum. 

Let us now consider a radiated Higgs boson in the three-body final state. The preceding argument for an unsuppressed $s$-wave still applies as the final state leptons need only recoil against a boson regardless of its scalar or vector nature. For massless final state fermions the only diagrams of Fig.~\ref{fig:feynmandiagrams} that contribute to the amplitude will be the middle two internal bremsstrahlung ones. It will be useful to look at the $\chi \chi \to H f \bar{f}$ amplitude in detail to illustrate explicitly how the helicity suppression is lifted, arguing analogously to the electroweak gauge boson case in Ref.~\cite{ciafalonietal2}. 

Labelling the initial state DM particles and final state fermions momenta by $p_1, p_2$ and  $p_3, p_4$ respectively, with the Higgs momentum denoted $p_5$, we may write for the process $\chi \chi \to H e^+ e^-$ the total amplitude corresponding to the internal bremsstrahlung diagrams as
\begin{eqnarray}
i \mathcal{M} - i \mathcal{M} _\text{exch.} &=&  y_{DM}^2 (-i \sqrt{2} \lambda_D v_\text{EW}) 	
\frac{1}{2}
\left[D_{24}D_{13}  \bar{v}(p_2) P_L \gamma^\mu u(p_1) - D_{23}D_{14}  \bar{v}(p_2) P_R \gamma^\mu u(p_1) \right] \nonumber	\\
&&\times \left[ \bar{u}(p_3) P_R \gamma_\mu v(p_4) \right]
\, ,
\label{eq:Hamplitude}
\end{eqnarray}
where the propagator factor $D_{ij}$ is defined as 
\begin{equation}
D_{ij} \equiv \frac{1}{(p_i - p_j)^2 - r_\pm m_\chi^2}  \,  .
\end{equation}
This expression is obtained after applying a Fierz transformation to the amplitude of Eq.~(\ref{eq:basicamplitude}) in Appendix~\ref{sec:calculation} in order to group the initial and final states into respective fermion bilinears. The amplitude for the process $\chi \chi \to H \nu_e \bar{\nu}_e$ can be obtained by the substitution $\lambda_D \to \lambda_D + \lambda_F$ and $r_\pm \to r_0$ in the above equations. 

The initial state bilinear of the current has a vector part proportional to 
\begin{equation*}
(D_{24}D_{13} - D_{23}D_{14}) \times \bar{v}(p_2)\gamma^\mu u(p_1)	\, ,
\end{equation*}
which is velocity suppressed since $D_{24}D_{13} - D_{23}D_{14} \sim \mathcal{O}(v)$ in the $v \ll 1$ limit, and for $r_\pm \gg 1$ this is $\sim \mathcal{O}(\frac{v}{{r_\pm}^3})\frac{1}{m_\chi^4}$. The axial vector part on the other hand is 
\begin{equation}
-(D_{24}D_{13} + D_{23}D_{14}) \times \bar{v}(p_2)\gamma_5\gamma^\mu u(p_1)		\, ,
\label{eq:propcoeff1}
\end{equation}
which has a coefficient proportional to $D_{24}D_{13} + D_{23}D_{14} \sim \mathcal{O}(\frac{1}{{r_\pm}^2})\frac{1}{m_\chi^4}$ in the large $r_\pm$ limit. We can then use the Gordon identity to rewrite this as
\begin{equation*}
\bar{v}(p_2)\gamma_5\gamma^\mu u(p_1) = \frac{(p_1+p_2)^\mu}{2m_\chi} \bar{v}(p_2)\gamma_5 u(p_1) 
+\frac{i}{2m_\chi}\bar{v}(p_2)\sigma^{\mu\nu}(p_2 - p_1)_\nu \gamma_5 u(p_1)	\,  .
\end{equation*}
The second term is also velocity suppressed since $(p_1 - p_2)^\mu \sim \mathcal{O}(v)m_\chi$, but the pseudoscalar term with the momentum sum $(p_1 + p_2)^\mu = (p_3 + p_4 + p_5)^\mu$ yields an un-suppressed $s$-wave contribution. Note that this momentum gets contracted into the final state fermion bilinear part of the current in Eq.~(\ref{eq:Hamplitude}), and for the two-body process we would have instead $(p_1 + p_2)^\mu = (p_3 + p_4)^\mu$ in a similar decomposition of the $2\to 2$ amplitude. Using the Dirac equation this is proportional to the final state fermion mass and hence is responsible for the helicity suppression. The inclusion of a third body with momentum $p_5^\mu$ in the final state is thus essential in opening up the $s$-wave.

\begin{figure}[h!]
\centering
\includegraphics[scale=1]{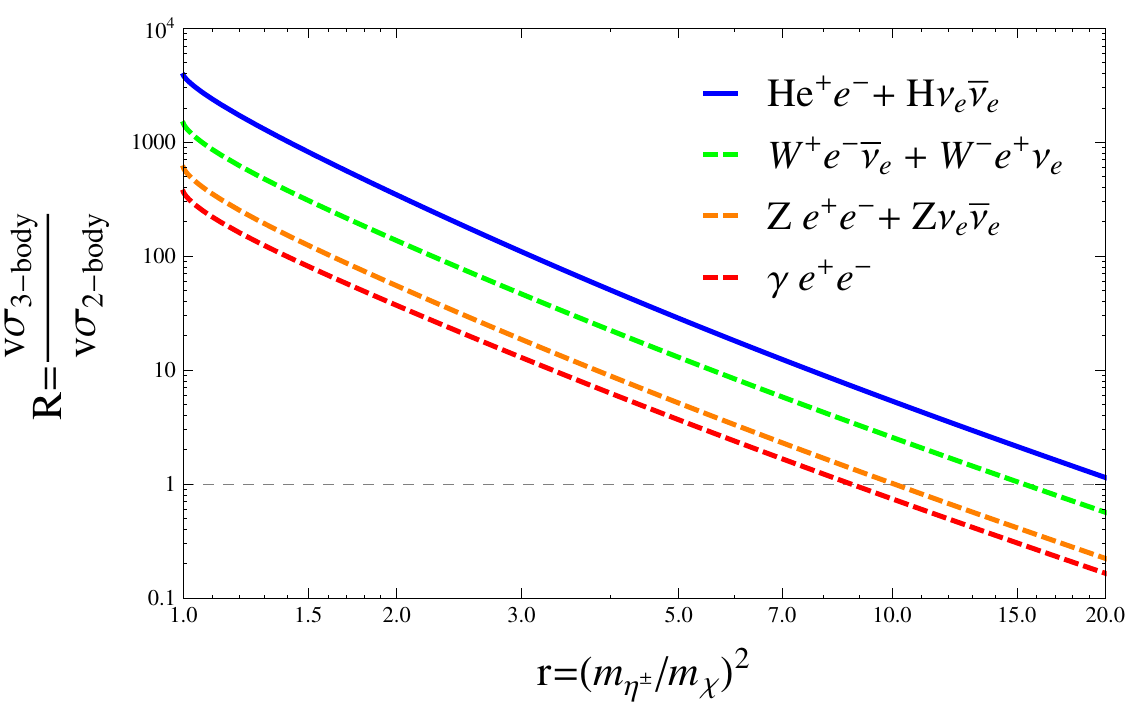}
\caption{\it 
DM annihilation cross-section to three-body final states $H, W^\pm, Z, \gamma$ by descending order of importance, normalised by the total two-body rate $\sigma v(\chi\chi \to e^+e^-) + \sigma v(\chi \chi \to \nu_e\bar{\nu}_e)$, as a function of various values of the mass of the mediating scalar $\eta$ parametrised by $r$. Here $m_\chi = 300$ GeV, $\lambda_D=1$, $\lambda_F=0$ and $v = 10^{-3}$. 
}
\label{fig:R_vs_r}
\end{figure}

The $s$-wave cross-section is obtained in Appendix \ref{sec:calculation} by integrating the squared amplitude over three-body phase space. Fig.~\ref{fig:R_vs_r} shows the result for the doublet model on a plot of the three- to two-body annihilation cross-section ratio $R$ as a function of varying $r \equiv (m_{\eta^\pm}/m_\chi)^2$, keeping $m_\chi$ fixed at 300 GeV, $\lambda_D = 1$, $\lambda_F = 0$ (corresponding to the degenerate scalar mass case $m_{\eta^\pm}=m_{\eta^0}$) and $v=10^{-3}$. The two-body cross-section in the ratio $R$ is defined as
\begin{equation}
v\sigma_\text{2-body} \equiv v\sigma(\chi\chi \to e^-e^+) + v\sigma(\chi\chi \to \nu_e\bar{\nu}_e)	\, .
\label{eq:2bodynorm}
\end{equation}
We have validated our results by comparing with those of Refs.~\cite{belletal1,ciafalonietal2} and find them to be consistent when the same conventions are taken into account. 

The dashed red, orange and green lines denote $\gamma, Z$ and $W^\pm$ bremsstrahlung respectively, by increasing order of strength, and we note that for the singlet model $W^\pm$ bremsstrahlung cannot occur. We see that the solid blue line representing Higgs-strahlung is in this case the dominant contribution. The ratios $R$ fall as expected when the scalar decouples with increasing $r$, but can become several orders of magnitude larger as the DM and mediator mass are increasingly degenerate. This scenario naturally occurs for example in neutralino-sfermion coannihilation regions of the MSSM parameter space, as mentioned earlier in Section~\ref{sec:darkmattermodel}.

\begin{figure}[h!]
\centering
\includegraphics[scale=1]{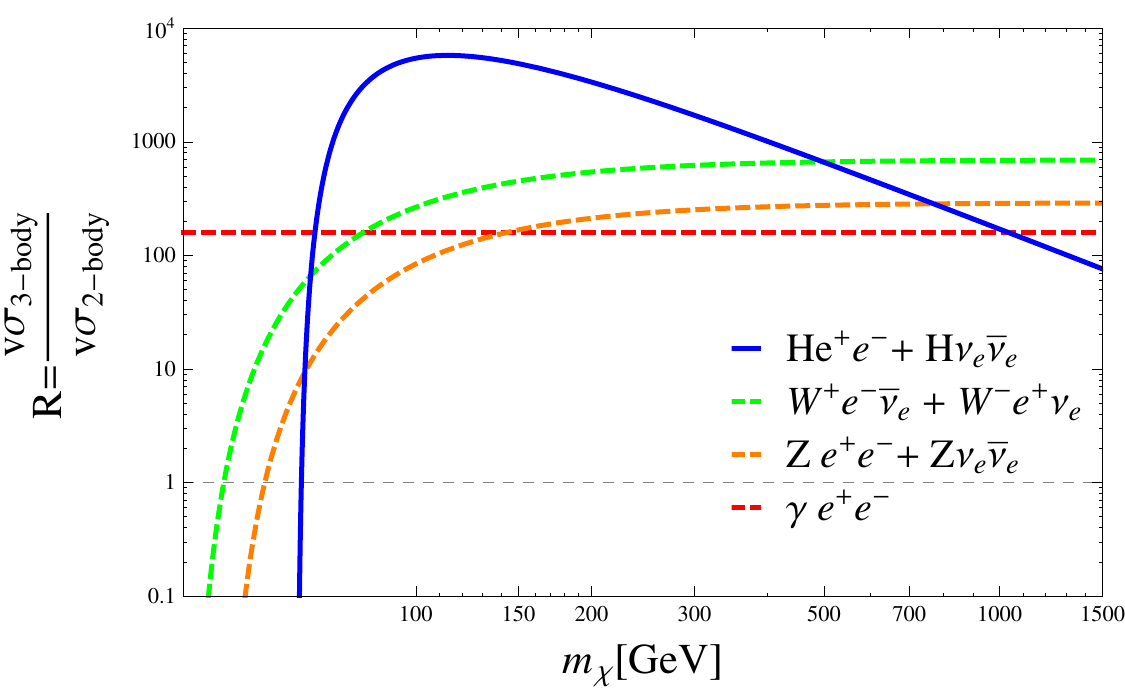}
\caption{\it 
DM annihilation cross-section to three-body final states $H, W^\pm, Z, \gamma$, normalised by the total two-body rate $\sigma v(\chi\chi \to e^-e^+) + \sigma v(\chi \chi \to \nu_e\bar{\nu}_e)$, as a function of the DM mass $m_\chi$ for $r=1.2$, $\lambda_D=1$, $\lambda_F=0$ and $v=10^{-3}$. 
}
\label{fig:R_vs_Mchi}
\end{figure}

Next in Fig.~\ref{fig:R_vs_Mchi} we look at the effect on the cross-section ratio $R$ of keeping the DM-mediator mass splitting parameter $r$ fixed to 1.2 while varying the DM mass $m_\chi$. The same line style scheme is used as previously. We see that Higgs-strahlung is important for DM mass below 1 TeV but drops faster with increasing $m_\chi$ than gauge boson bremsstrahlung. This is expected from the dimensionful coupling $\sim v_\text{EW}$ of the Higgs to the $\eta^\pm$ and $\eta^0$ scalars which leads to an additional $1/m_\chi^2$ dependence.

\begin{figure}[h!]
\centering
\includegraphics[scale=1]{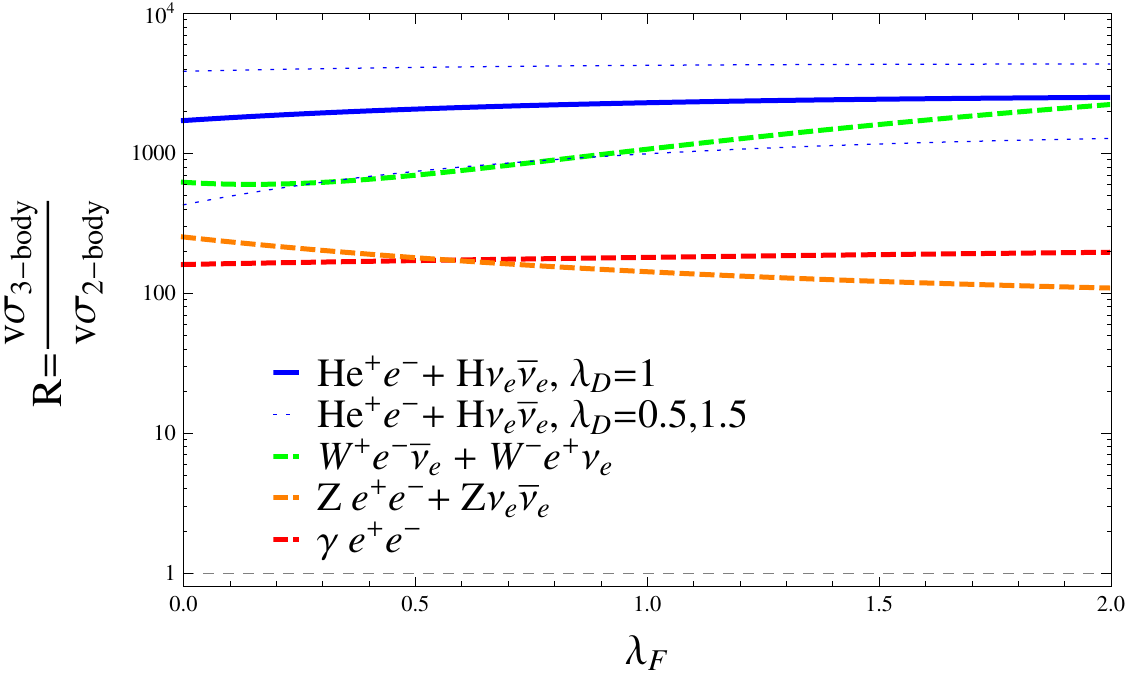}
\caption{\it 
DM annihilation cross-section to three-body final states $H, W^\pm, Z, \gamma$, normalised by the total two-body rate $\sigma v(\chi\chi \to e^+e^-) + \sigma v(\chi \chi \to \nu_e\bar{\nu}_e)$, as a function of the scalar mass degeneracy parameter $\lambda_F$ for $r_\pm = 1.2, m_\chi = 300$ GeV and $v=10^{-3}$. $\lambda_D=1$ for the solid blue line, with the dotted blue lines denoting the Higgs-strahlung cross-section range when varying $\lambda_D$ from 0.5 to 1.5.
}
\label{fig:R_vs_lambdaF}
\end{figure}

In Ref.~\cite{garny1} the importance of contributions from the longitudinal component of the $W^\pm$ when $m_{\eta^\pm} \neq m_{\eta^0}$ was highlighted for the doublet model. Fig.~\ref{fig:R_vs_lambdaF} compares the annihilation cross-section ratio $R$ for $W^\pm$ bremsstrahlung (dashed green line) for varying values of the scalar mass degeneracy parameter $\lambda_F$ against the cross-section from the Higgs (solid blue line) and $Z,\gamma$ (dashed orange, red lines). The parameters used are $m_\chi = 300$ GeV, $r_\pm = 1.2$, $\lambda_D = 1$ in solid blue and $\lambda_D = 0.5\,(1.5)$ for the lower (upper) dotted blue lines. 

We see that the $W^\pm$ contribution grows as the longitudinal component increases with large $\lambda_F$. This component is proportional to the scalar mass splitting since it originates from the Goldstone boson $G^\pm$ coupling to the $\eta^0$ and $\eta^\pm$ fields in the Feynman gauge, whose coefficient can be written as $\lambda_F v_\text{EW} = \frac{1}{v_\text{EW}}(m_{\eta^0}^2 - m_{\eta^\pm}^2)$ after electroweak symmetry breaking. This is of the same form as the Higgs-strahlung coupling $\sqrt{2}\lambda_D v_\text{EW}$ and  $\sqrt{2}(\lambda_D+\lambda_F)v_\text{EW}$ from the Lagrangian terms for $H \eta^+ {\eta^+}^*$ and $H\eta^0{\eta^0}^*$ respectively. 

Note that the two-body normalisation of $R$ defined in Eq.~(\ref{eq:2bodynorm}) also depends on $\lambda_F$ through the mass of the neutral scalar in the propagator which suppresses the two-body annihilation rate. The decrease in the two-body cross-section is reflected in the slight increase of the dashed red line, since photon bremsstrahlung is independent of $\lambda_F$. On the other hand the $\chi\chi \to Z\nu_e\bar{\nu}_e$ amplitude also has a propagator with a dependence on the mass of the $\eta^0$ that suppresses the cross-section as $\lambda_F$ becomes large, and unlike the $W^\pm$ there is no enhancement from the longitudinal component.

\section{Energy Spectra of Final States}
\label{sec:energyspectra}

Indirect detection experiments search for DM through the spectrum of stable final states after its self-annihilation, and the inclusion of a radiated Higgs will affect this expected cosmic ray flux. In this section we investigate the energy spectrum of the stable SM particles after the Higgs decay. The most promising channels to disentangle a signal from astrophysical background are the photon, neutrino, antiproton and positron final states, which we will focus on here. 

\begin{figure}[h!]
\centering
\includegraphics[scale=0.69]{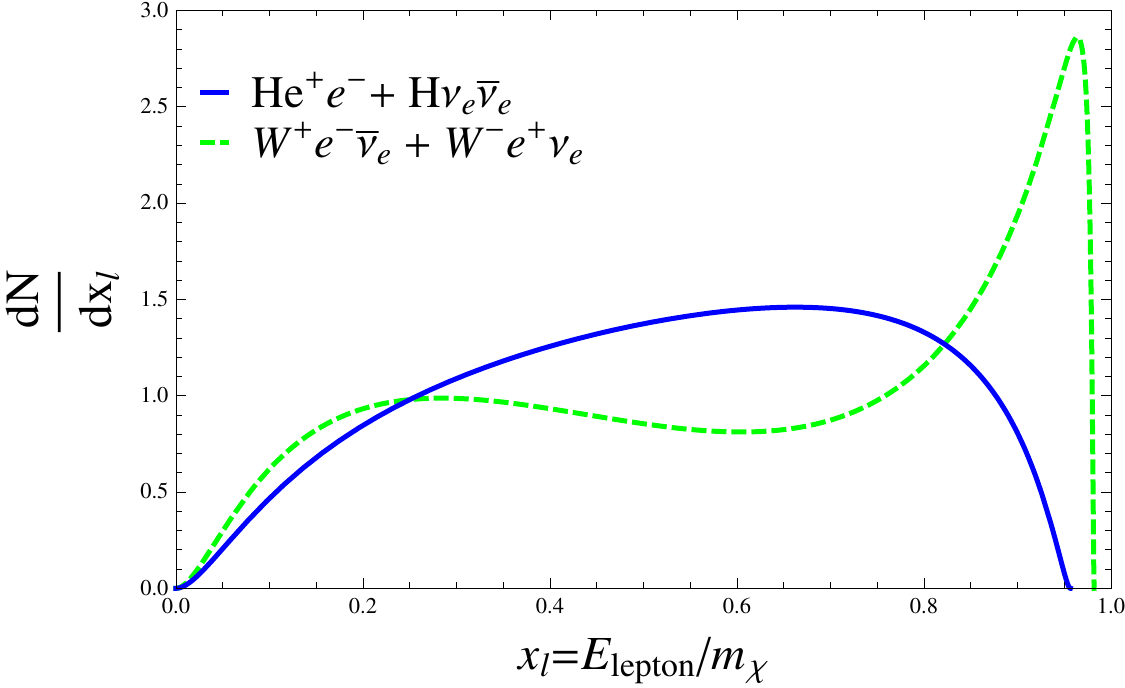}
\includegraphics[scale=0.69]{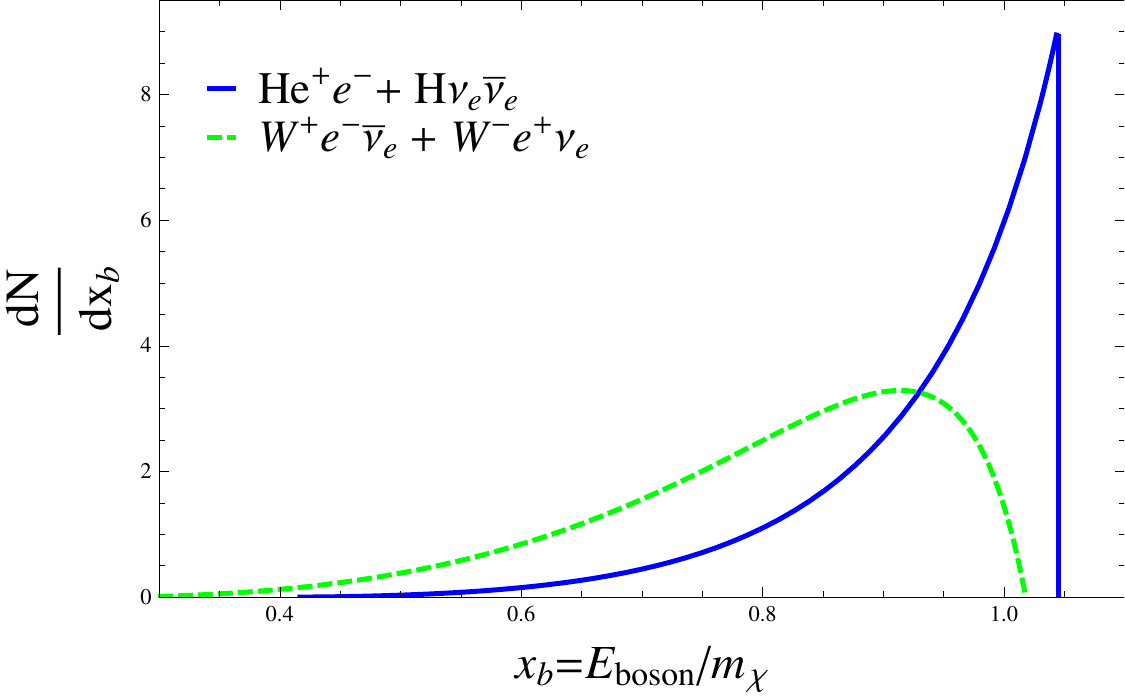}
\caption{\it 
Normalised energy distribution of the lepton (boson) on the left (right) originating from the hard process in DM annihilation for $m_\chi=300$ GeV, $r_\pm = r_0 = 1.2$, $\lambda_D=1$ and $\lambda_F=0$. Solid blue (dashed green) lines denote Higgs-strahlung ($W^\pm$-strahlung) processes.
}
\label{fig:energydistribution}
\end{figure}

We start with the energy spectrum of the lepton, antilepton and boson originating from the hard process of the DM annihilation. Fig.~\ref{fig:energydistribution} displays on the left (right) the differential energy distribution 
\begin{equation*}
\frac{dN}{dx} = \frac{1}{v\sigma(\chi\chi \to Bf\bar{f})}\frac{vd\sigma(\chi\chi \to B f\bar{f})}{dx}
\end{equation*}
as a function of $x \equiv E / m_\chi$ of the lepton (boson) produced in the Higgs- and $W^\pm$-bremsstrahlung processes, denoted by the solid blue and dashed green lines respectively. These are obtained from the analytical expression in Eq.~(\ref{eq:dsigma}) of Appendix \ref{sec:calculation} with $m_\chi =300$ GeV, $r_\pm = r_0 = 1.2$, $\lambda_D=1$ and $\lambda_F=0$. We will use these representative values throughout this section. 

The subsequent decay and fragmentation of the radiated bosons $B = H, Z, W^\pm$ is handled in \textsc{Pythia} 8.176~\cite{pythia}. We have written our own Monte-Carlo (MC) that generates events for each three-body process $\chi \chi \to B f \bar{f}$ by randomly sampling the volume of the double-differential cross-section over the kinematic phase space. These are then passed to \textsc{Pythia} in order to simulate the subsequent showering into stable SM particles. We have checked that the MC reproduces the distributions of Fig.~\ref{fig:energydistribution} when the boson decay is switched off, and validated the results after decay by comparing with Ref.~\cite{garny1}. 

\begin{figure}[h!]
\centering
\includegraphics[scale=0.38]{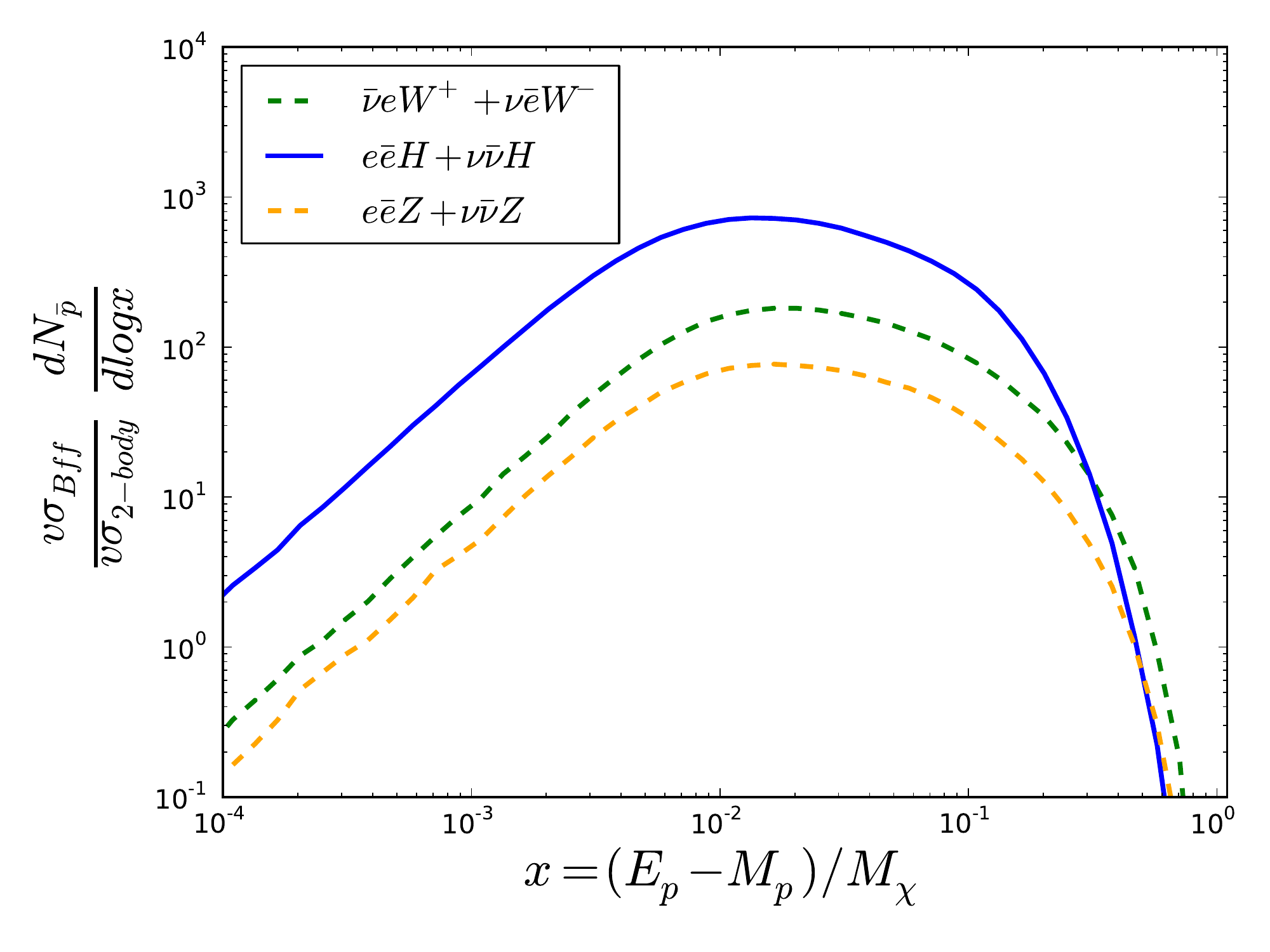}
\includegraphics[scale=0.38]{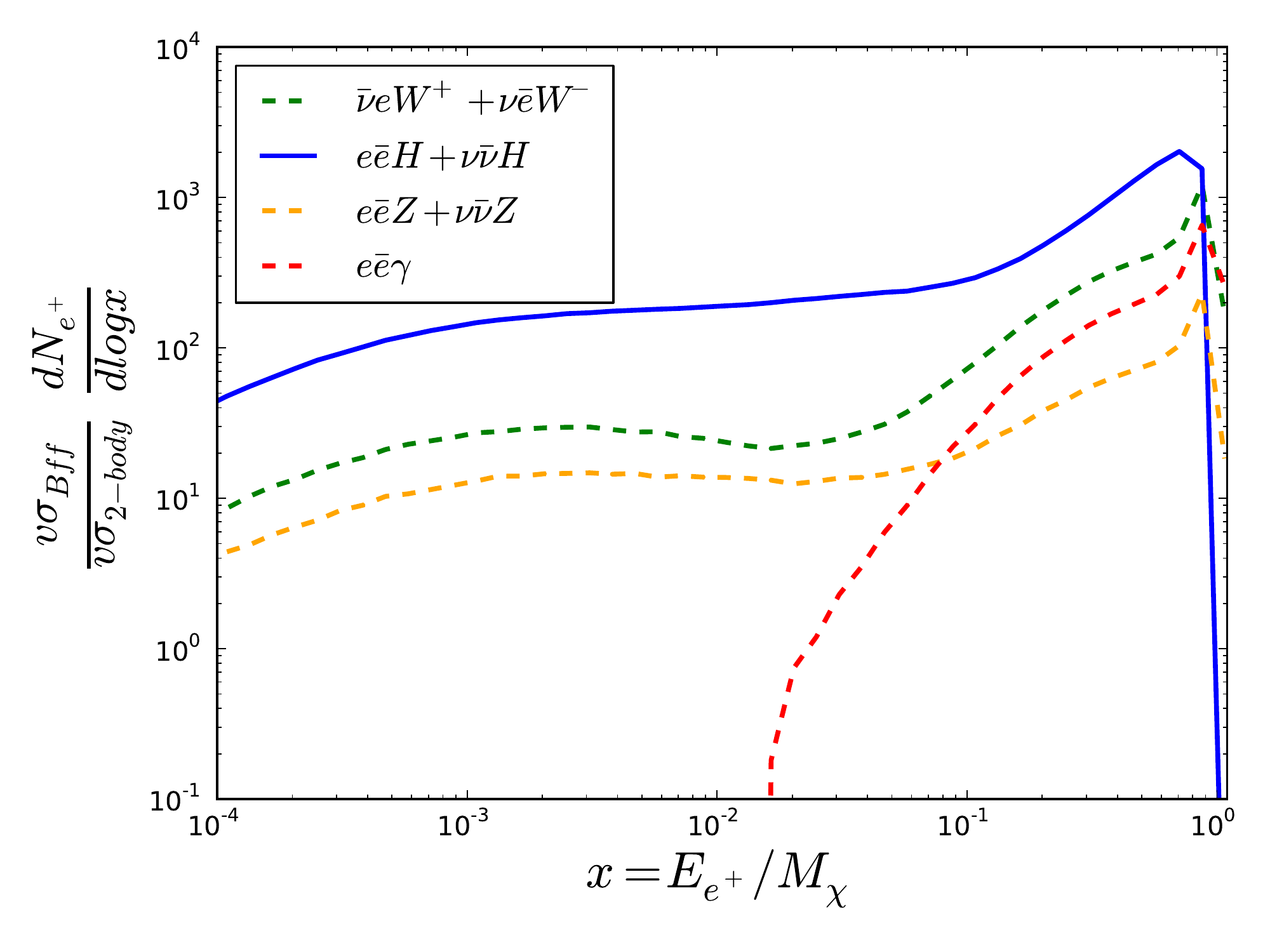} \\
\includegraphics[scale=0.38]{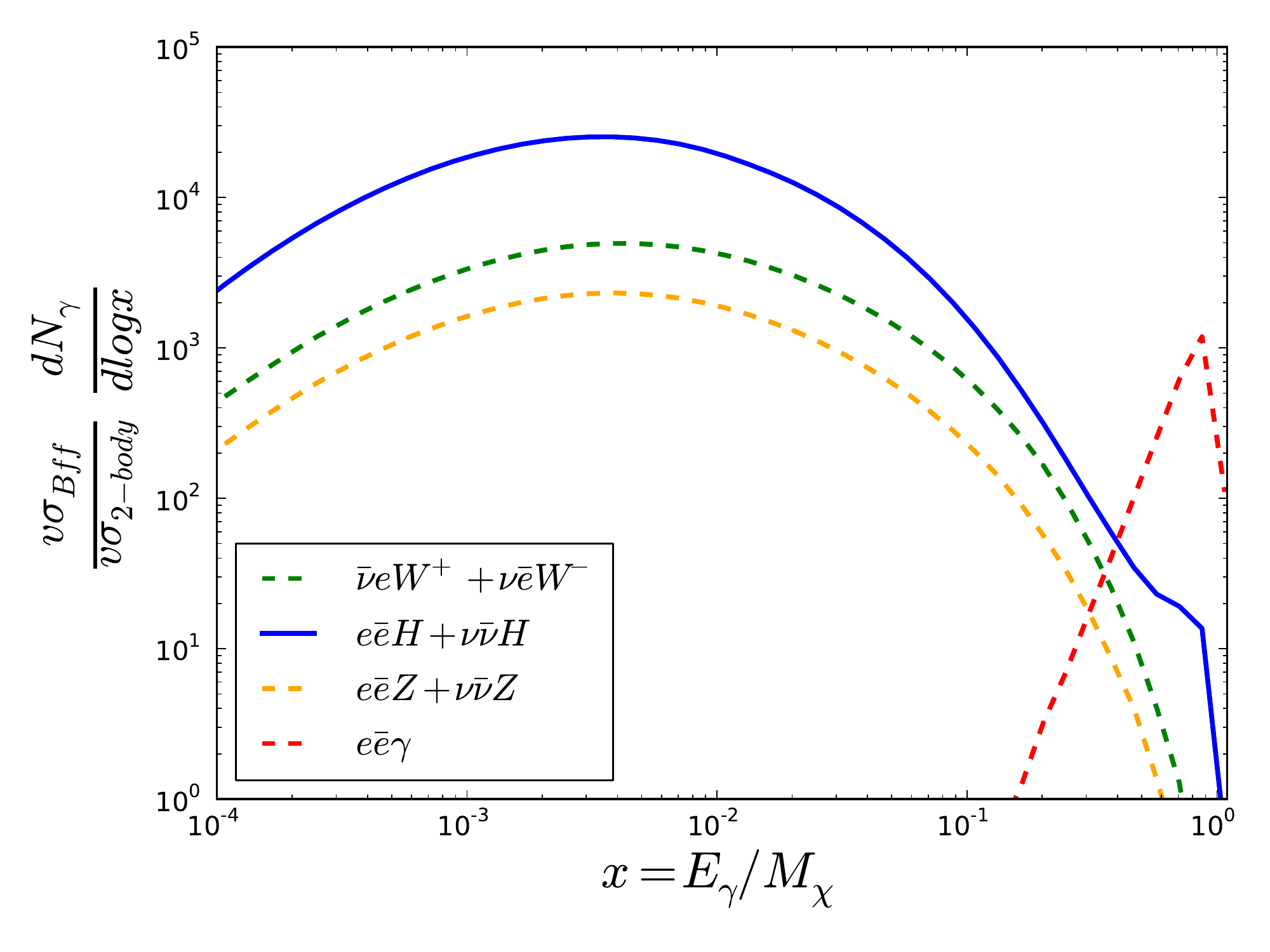}
\includegraphics[scale=0.38]{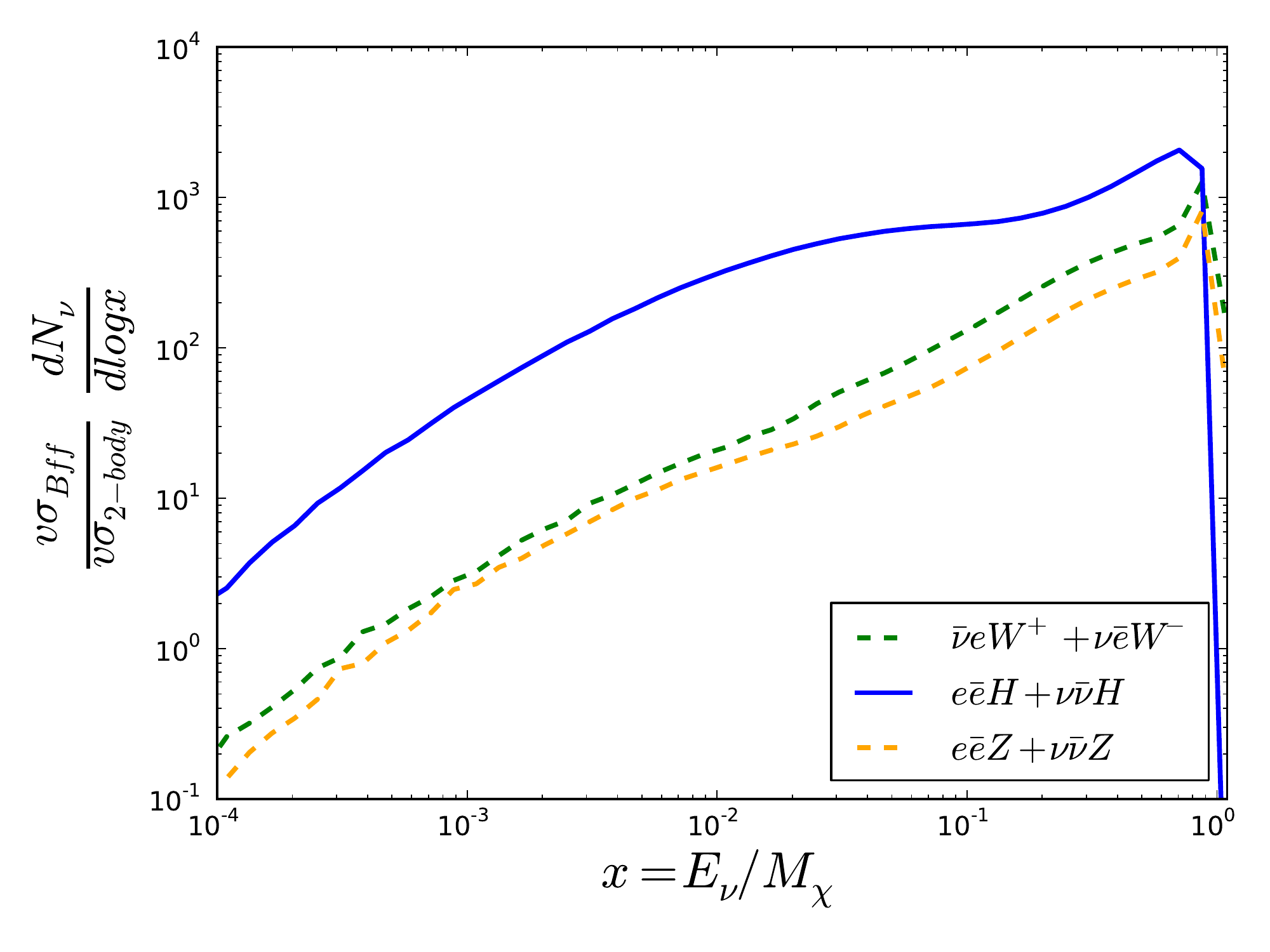}
\caption{\it 
Normalised kinetic energy distribution of final stable particles after showering for $m_\chi=300$ GeV, $r_\pm = r_0 = 1.2,\lambda_D=1$ and $\lambda_F=0$. The antiproton ($\bar{p}$), positron ($e^+$), neutrino ($\nu_e + \nu_\mu + \nu_\tau$) and photon ($\gamma$) final states are displayed clockwise from the top left. Each channel $W^\pm,Z,\gamma$ and $H$ is shown separately by dashed green, orange, red and solid blue lines respectively.
}
\label{fig:energyeachchannel}
\end{figure}

Using this setup a total of $9\times10^6$ events were generated with the relative number of events for each channel,
\begin{equation*} 
B_i f \bar{f} = \{W^+ e^- \bar{\nu}_e + W^- e^+ \nu_e, Ze^+ e^- + Z\nu_e\bar{\nu}_e, He^+ e^- + H\nu_e \bar{\nu}_e, \gamma e^+ e^-\}	\, ,
\end{equation*}
proportional to their cross-sections. In Fig.~\ref{fig:energyeachchannel} we fit the numerical results and plot the individual normalised spectrum
\begin{equation*}
\frac{v\sigma_{B_i f\bar{f}}}{v\sigma_\text{2-body}} \frac{dN^i_j}{d\text{log}x_j} 
= \frac{1}{v\sigma_\text{2-body}} \frac{vd\sigma(\chi\chi \to B_i f\bar{f} \to p_j + \text{...})}{d\text{log}x_j} 	
\, , \, x_j = \frac{E^j_\text{kinetic} }{m_\chi}	\, ,
\end{equation*}
for each channel $i$ separately. The flux originating from $W^\pm,Z, \gamma$ and $H$ bremsstrahlung are denoted by dashed green, orange, red and solid blue lines respectively. The final stable particles $p_j = \bar{p}, e^+, (\nu_e + \nu_\mu + \nu_\tau), \gamma$ labelled by $j$ are displayed clockwise starting from the anti-proton spectrum on the top left. The photon channel only contributes to the gamma spectrum, with the famous bump at high energy, and to the positron spectrum from the primary final state leptons. The electroweak bremsstrahlung on the other hand opens up the hadronic decay to antiprotons despite our leptophilic model, with subsequent showers generating a low-energy tail of additional leptons and photons. In addition to these well-known processes, we see a significant addition to the spectrum from Higgs-strahlung. 

\begin{figure}[h!]
\centering
\includegraphics[scale=0.7]{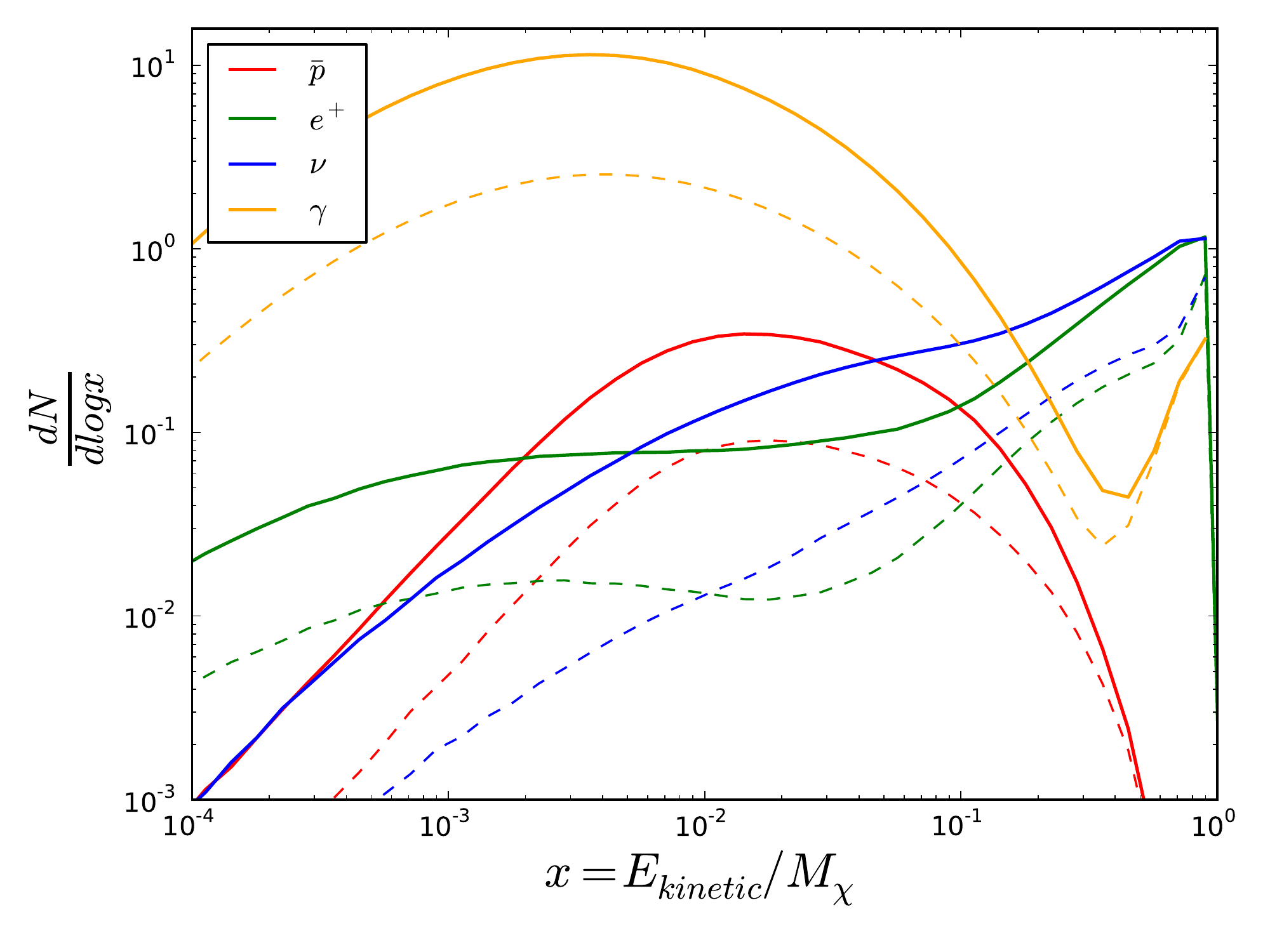}
\caption{\it 
Normalised kinetic energy distribution of final stable particles after showering with all bremsstrahlung channels combined, using $m_\chi=300$ GeV, $r_\pm = r_0 = 1.2, \lambda_D=1$ and $\lambda_F=0$. The antiproton ($\bar{p}$), positron ($e^+$), neutrino ($\nu_e + \nu_\mu + \nu_\tau$) and photon ($\gamma$) final states are represented by red, green, blue and yellow lines respectively. Solid lines include all bremsstrahlung channels while dashed lines are electroweak and photon bremsstrahlung only without Higgs-strahlung.
}
\label{fig:energycombined}
\end{figure}

Combining all the channels together yields the final energy spectrum at the annihilation source. In Fig.~\ref{fig:energycombined} we plot the distribution 
\begin{equation*}
\frac{dN_j}{d\text{log}x_j} \equiv \sum_i \frac{v\sigma_{B_i f\bar{f}} }{v\sigma_\text{all channels}} \frac{dN^i_j}{d\text{log}x_j} 
\end{equation*}
for each final stable state $j$. Here the stable particles $p_j = \bar{p}, e^+, (\nu_e + \nu_\mu + \nu_\tau), \gamma$ are denoted by red, green, blue and yellow lines. The solid lines include all contributions from electroweak, photon and Higgs bremsstrahlung while dashed lines represent the electroweak and photon channels only. 

The distribution calculated here must be propagated from the annihilation point to the Earth, with various astrophysical uncertainties and solar modulation taken into account, in order to obtain the final flux of cosmic rays measured by experiments. Any features in the positron spectrum would be washed out by this process while the antiproton spectrum would be less affected. The neutrino and gamma ray features are essentially expected to be preserved. A full simulation would take us beyond the scope of this work as our aim is not to place exclusion limits but to highlight the importance of Higgs-strahlung contributions.

\section{Conclusion}
\label{sec:conclusion}

We have calculated the effects of including a radiated Higgs in Majorana DM annihilation to two leptons and found this to dominate over photon and electroweak bremsstrahlung for $m_\chi \lesssim 1$ TeV and $\lambda_D \sim \mathcal{O}(1)$. This holds over the usual range $m_\eta \lesssim 4 m_\chi$ in which annihilation to three-body final states is larger than annihilation to two leptons. The Higgs coupling to the mediating scalars is parametrically similar to the longitudinal $W^\pm$ which can also take part in bremsstrahlung processes, but unlike this latter case the Higgs coupling does not vanish in the limit of equal charged and neutral scalar mass. We also note that for models in which the mediating scalar is an SU$(2)_L$ singlet the Higgs-strahlung contribution remains while $W^\pm$ bremsstrahlung no longer plays a role.  

Taking into account Higgs-strahlung we find that the decay and showering of final state particles yields a significantly higher flux of stable SM particles than with only photon and electroweak bremsstrahlung. Given the generic nature of this process we argue that it should be included in any realistic exclusion limits based on antiproton or positron signatures, and in searches for neutrinos from solar DM annihilation which are sensitive to the same DM mass range $\lesssim 1$ TeV in which Higgs-strahlung is significant.

\section*{Acknowledgements}
The authors thank John Ellis, Keith Olive and Michihisa Takeuchi for helpful comments and discussions, as well as the hospitality of CERN where part of this work was carried out. F.~L. was supported by the London Centre for Terauniverse Studies (LCTS) using funding from the European Research Council via the Advanced Investigator Grant 267352. T.~Y. was supported by a KCL Graduate Teaching Assistantship.

\appendix

\section{Calculation of Bremsstrahlung Cross-Section}
\label{sec:calculation} 

The amplitude for the $\chi\chi \to Hf\bar{f}$ process for massless final state fermions may be written as 
\begin{align}
i \mathcal{M}_\text{tot.} &= i \mathcal{M} - i \mathcal{M}_\text{exch.}	\, , \nonumber \\
i \mathcal{M} &= y_{DM}^2 (-i \sqrt{2} \lambda v_\text{EW}) \frac{[\bar{v}(p_2)P_L v(p_4)] \cdot [\bar{u}(p_3)P_R u(p_1)]}{\left[(p_2-p_4)^2-m_\eta^2 \right] \left[(p_1-p_3)^2-m_\eta^2\right]}	\, , \nonumber \\
i \mathcal{M}_\text{exch.} &= y_{DM}^2 (-i \sqrt{2} \lambda v_\text{EW}) \frac{[\bar{v}(p_1)P_L v(p_4)] \cdot [\bar{u}(p_3)P_R u(p_2)]}{\left[(p_1-p_4)^2-m_\eta^2\right] \left[(p_2-p_3)^2-m_\eta^2\right]}	\, ,
\label{eq:basicamplitude}
\end{align}
where $p_1,p_2$ and $p_3,p_4$ label the initial and final state fermion four-momenta, and $p_5$ is the Higgs four-momentum. For the final states $H f\bar{f} = H e^-e^+, H \nu_e\bar{\nu}_e$ we have $m_\eta = m_{\eta^\pm}, m_{\eta^0}$ and $\lambda = \lambda_D, \lambda_D + \lambda_F$ respectively. This can then be Fierz-transformed into the form of Eq.~(\ref{eq:Hamplitude}). 

The annihilation cross-section is, averaging over initial state spins and summing over final state spins, 
\begin{equation*}
\sigma v = \frac{1}{2s}\int d\Phi_\text{3-body} \frac{1}{4}\ssum_\text{spin}|\mathcal{M}_\text{tot.}|^2 	\,  .
\end{equation*}
In the small velocity limit certain redundant choice of angles may be integrated out so the three-body phase space integral can then be decomposed as 
\begin{equation*}
\int d\Phi_\text{3-body}(p_3,p_4,p_5) = \int_{m_H^2}^s \frac{dq^2}{2\pi}\int_{-1}^1\frac{d\text{cos}\theta}{2} \frac{\bar{\beta}(p_4,p_5)}{8\pi} \frac{\bar{\beta}(q,p_3)}{8\pi}	\, ,
\end{equation*}
where $q = p_4 + p_5$ and $\bar{\beta}$ is defined as
\begin{equation*}
\bar{\beta}(p_A,p_B) \equiv \sqrt{1- \frac{2(p_A^2 + p_B^2)}{(p_A+p_B)^2} + \frac{(p_A^2 - p_B^2)^2}{(p_A+p_B)^4}} \,  . 
\end{equation*}
The differential cross-section for $v \ll 1$ in terms of dimensionless variables $r_H = (m_H/ 2 m_\chi)^2, r_q = q^2/s$ and $r = (m_\eta/m_\chi)^2$ is found to be 
\begin{equation}      
\frac{v d \sigma}{d r_q} = \frac{\lambda^2 v_\text{EW}^2 y_{DM}^4}{256 \pi ^3 m_\chi^4}  \frac{r_q (r-2r_H+2r_q-1) \, \ln \left[\frac{r_q (r-2r_H+2r_q-1)}{r_q r -2
  r_H+ r_q}\right]+2 (r_q-1) (r_H- r_q)}{ (r-2 r_q+1)^2 (r-2r_H+2
  r_q-1)}     \, .
\label{eq:dsigma}
\end{equation}
Integrating this in the large $r$ limit we obtain
\begin{equation}
\sigma v |_{r\to \infty} = \frac{\lambda^2 v_\text{EW}^2 y_{DM}^4}{1536 \pi ^3 m_\chi^4 \, r^4} \left[ 1- 8 r_H + 8 r_H^3 -r_H^4 - 12 r_H^2 \ln (r_H)  \right] 
\, ,
\label{eq:large_r_limit}
\end{equation}
which is of the same form as the longitudinal $W^\pm$ bremsstrahlung cross-section given by Eq.~(A.5) in Ref.~\cite{garny1}. It is also interesting to integrate Eq.~(\ref{eq:dsigma}) in the limit $r \to 1$, where the annihilation enhancement is largest. This gives
\begin{equation}
\sigma v |_{r\to 1} = \frac{\lambda^2 v_\text{EW}^2 y_{DM}^4}{1024 \pi ^3 m_\chi^4} \left[\text{Li}_2(1- r_H)+\frac{r_H \ln (r_H)}{r_H-1}-1 \right]
\, .
\label{eq:small_r_limit}
\end{equation}

We have also calculated in this way the corresponding expressions for $W^\pm,Z$ and $\gamma$ bremsstrahlung. These are available for example in Refs.~\cite{originalbremsstrahlung,belletal1,ciafalonietal2}.


$ $

\end{document}